\documentclass[nofootinbib,showpacs,showkeys]{revtex4-2}
\usepackage{graphicx,color}
\usepackage{amssymb}
\usepackage{amsmath}
\usepackage{bm}
\usepackage{tensor} 
\usepackage{lipsum}
\usepackage{latexsym}
\usepackage{dcolumn}
\usepackage{float}
\usepackage{hyperref}
\begin{document}

\title{Skewness as a test of dark energy perturbations}
\author{Raquel Emy Fazolo}%
 \email{raquel.fazolo@edu.ufes.br}
\affiliation{ PPGCosmo, Universidade Federal do Esp\'irito Santo, 29075-910, Vit\'oria, ES, Brazil}
\author{Luca Amendola}
\email{l.amendola@thphys.uni-heidelberg.de}
\affiliation{Institute of Theoretical Physics, Philosophenweg 16, Heidelberg University, 69120, Heidelberg, Germany}
\author{Hermano Velten}%
 \email{hermano.velten@ufop.edu.br}
\affiliation{Departamento de F\'isica, Universidade Federal de Ouro Preto (UFOP), Campus Morro do Cruzeiro, 35400-000, Ouro Preto-MG Brazil}
\date{\today}

\begin{abstract}
We investigate the role played by dark energy perturbations in the skewness $S_3$ of large-scale matter distribution. We consider a two-fluid universe composed by matter and dark energy, with perturbations in both components, and we estimate numerically the skewness of the matter density field as a function of the dark energy parameters. We characterize today's $S_3$ value for quintessence and phantom dark energy cosmologies as well as its dependence on the matter density parameter $\Omega_{m0}$ and  the dark energy sound speed $c^2_s$ with accurate numerical fitting. These fits can be used to test cosmology against future high quality data on large scale structure. 
\end{abstract}

\maketitle

\section{\label{sec:Intro}Introduction}

Observations show that today's universe is composed mostly of dark energy ($\approx 70\% $), a substance responsible for the current accelerated expansion which also, under normal circumstances, slows down the clustering  of the universe's matter content. As well known, the simplest form of dark energy (DE) is a constant $\Lambda$ in Einstein's general relativity acting only at expanding background level. A fluid model for dark energy is however also a valid description allowing for a time evolution of its energy density. The implementation of the fluid formalism demands the definition of an equation of state parameter (EoS), $w=P/\rho$, defined as the ratio of the fluid's pressure $P$ and its energy density $\rho$. For the dark energy case, even though a time evolving EoS parameter provides a vast collection of possible dynamical scenarios, a constant $w$ is enough to produce a rich phenomenology of dark energy models.

In a first approximation, the late-time universe can be described as a homogeneous and isotropic expanding background filled with matter and dark energy. Aspects of large scale structure are studied mainly via the definition of the matter density contrast $\delta \rho_{m}$ and dark energy perturbations are usually neglected or are shown not to be able to contribute to the linear regime of structure formation \cite{Amendola:2003wa}. The latter assumption is formally obeyed by a cosmological constant since it does not cluster and, to a good approximation, by time-evolving dark energy models with speed of sound close to unity \cite{Abramo:2007iu,Abramo:2008ip,Mota:2004pa,Batista:2017lwf,Fasiello:2016qpn,Sefusatti:2011cm,Endo:2018xhx}. On the other hand, dark energy perturbations in nonadiabatic dark energy models are severely constrained by observations \cite{Zimdahl:2019pqg}.

While the role played by DE perturbations at first order has been  widely studied in the literature \cite{Rajvanshi:2018xhf,Pace:2011kb,Abramo:2009ne,Mehrabi:2015hva,Pace:2017qxv}, dark energy effects at higher clustering perturbative orders are still poorly explored: for example, and more specifically for the goal of this work, the effects of dark energy perturbation on the skewness ($S_{3}$) of the matter density. The skewness, defined as the normalized third order moment of the counts-in-cells statistics, can be used to evaluate asymmetric features in the probability distribution function  of the matter clustering field. A finite skewness indicates the existence of a non-vanishing balance between clustering and voids in the large scale structure distribution. Indeed, it is well known that the cosmological gravitational clustering process yields to a skewed large scale structure \cite{Bernardeau:1993qu,Bernardeau:2001qr}. In a fully matter-dominated universe (the Einstein-de Sitter model), one finds the well-known result $S_3\simeq34/7 \simeq 4.857$ \cite{1980lssu.book.....P,Bernardeau:2001qr}, to be corrected for finite-volume effects.  

In previous work \cite{Velten:2019edo}, by following the treatment used in Ref. \cite{Reis:2004hm}, dark energy effects on $S_3$ have been studied by modelling the entire cosmic substratum, i.e., matter and dark energy, as a single fluid. By encoding DE effects into an effective equation of state and an effective speed of sound for the total matter fluid, a significant increase of the skewness value ($S_3 \sim 15$) was found for any dynamical dark energy model while skewness for $\Lambda$CDM model remains close to $34/7$. This result, while still employing a simplistic treatment, has showed that the skewness $S_3$ is a promising test to identify cosmic signatures of DE perturbations. While being an interesting cosmological observable, the actual $S_3$ value is still poorly constrained by the current observations \cite{Wolk:2013db} but in the near future, with surveys like Euclid \cite{Amendola:2016saw,EUCLID:2011zbd}, a reliable map the three-dimensional matter distribution of the universe can provide an accurate measurement of $S_3$. On the other hand, large scale N-body simulations can be used to investigate skewness as done recently in Ref. \cite{Einasto:2020sqi}.

Our aim with this work is to continue the search for dark energy signatures on the skewness of matter distribution. Now, going beyond the analysis performed in Ref. \cite{Velten:2019edo}, we fully take into account fluctuations in both cosmological fluids using the   perturbed equations up to second order for a universe composed by matter and dark energy. The dark energy density contrast acts as an independent  source of the matter fluctuations.

The main product of this paper is embodied in a series of numerical fitting formulas for the $S_3$ value calculated as a function of $\Omega_{m0}$ and $w_{de}$, with different choices of the dark energy sound speed $c_s^2$. These fits  might be useful when forecoming high-precision data from large scale structure surveys will be able to measure $S_3$ to a percent precision.
We expand over previous results in several ways: {\it a}) we include the dark energy perturbations; {\it b}) we simultaneously fit for $\Omega_{m0}$ and $w_{de}$; {\it c}) we adopt different values for $c_s^2$. The main equations of our work are presented in the next section. In the third section we provide the numerical analysis and the fitting formulas for $S_3$. We conclude in the final section. 

\section{\label{sec:Dynamics}Background and Perturbed Dynamics}

Since we focus on the late time cosmological dynamics (where the radiation contribution is negligible), we consider a non-interacting two-fluid model composed by pressureless matter $P_m=0$ (subscript $m$) and dark energy with pressure  $P_{de}=w_{de}\rho_{de}$ (subscript $de$). In a general-relativistic based description of the gravitational interaction sourced by matter and dark energy, Einstein's equations take the form
\begin{equation}
    R_{\mu\nu}-\frac{1}{2}g_{\mu\nu}\mathcal{R}=8\pi G\tensor{T}{_{(m)\mu\nu}}+8\pi G\tensor{T}{_{(de)\mu\nu}},
\end{equation}
being $R_{\mu\nu}$ the Ricci tensor and $\mathcal{R}=g^{\mu\nu}R_{\mu\nu}$ the Ricci scalar.
Both  components are described by a perfect fluid energy momentum tensor.  Using the background expansion rate of a flat, homogeneous and isotropic universe - the Friedmann-Lemaître-Robertson-Walker (FLRW) metric - one finds
\begin{equation}\label{H}
    \frac{H^{2}}{H^{2}_{0}}=\Omega_{m0}a^{-3}+\Omega_{de0}\,e^{-3\int da\frac{1+w_{de}}{a}}.
\end{equation}
where $H=\dot{a}/a$ is the Hubble expansion rate parameter, a dot represents the derivative with respect to cosmic time, and the  scale factor is normalized to the present value $a_0=1$. Since we assume a flat universe, we have $\Omega_{de0}=1-\Omega_{m0}$. 

Let us now focus on the cosmological perturbation analysis needed to calculated the skewness of matter distribution. Since on sub-horizon scales Newtonian physics  describes fairly well the cosmological expansion and its perturbations, we restrict our formalism to a post-Newtonian approach, or neo-Newtonian, as e.g. in \cite{Lima:1996at, Reis:2003fs}. 
Then, within the fluid limit, the cosmological dynamics can be recast in the form of a hydrodynamical system with velocity field given by the Hubble's law $\vec{u}=H \vec{r}$. In this formalism energy conservation is expressed in terms of the following continuity equation \cite{Lima:1996at}
\begin{equation}\label{cont}
    \left(\frac{\partial \rho}{\partial t}\right)_{r}+\vec{\nabla}_{r}\cdot(\rho\vec{u})+P\vec{\nabla}_{r}\cdot\vec{u}=0.
\end{equation}

The momentum conservation of a fluid moving under the influence of the gravitational field $\phi$ is represented by the Euler equation 
\begin{equation}\label{Euler}
    \left(\frac{\partial \vec{u}}{\partial t}\right)_{r}+(\vec{u}\cdot\vec{\nabla}_{r})\vec{u}=-\vec{\nabla}_{r}\phi-(\rho+P)^{-1}\vec{\nabla}_{r}P.
\end{equation}
It is  worth noting that we are going to neglect shear and vorticity in the Euler equation, a widely-employed approximation that allows our resulting set of equations to be closed.

Each individual fluid obeys  equations (\ref{cont}) and (\ref{Euler}). In the case of a $N$-fluid system, the entire dynamics will be given by $2N$ equations sourced by the function $\phi$, i.e., $2N+1$ variables. Thus, the system of equations will be closed by considering the Poisson equation (in the neo-Newtonian cosmology) relating the gravitational field to the contribution of the components such that
\begin{equation}
    \nabla^{2}\phi=4\pi G(\rho+3P).
\end{equation}

For a system composed by $N=2$ components, i.e.  matter and dark energy, we end up with a set of 5 equations. This post-Newtonian approach allows one to obtain the relativistic Friedmann Eq. (\ref{H}) with the identification $\vec{u}=H \vec{r}$.

Now, in order to assess the statistical moments of the matter density field, we introduce small perturbations around the background quantities, i.e.,
\begin{align}
    \rho&=\rho_{0}+\delta\rho, \\ \nonumber
    P&=P_{0}+\delta P, \\ \nonumber
    \phi&=\phi_{0}+\varphi, \\ \nonumber
    \vec{u}&=\vec{u}_{0}+\vec{v}.
\label{perturb}
\end{align}
 It is also convenient to adopt the comoving coordinates $\vec x$ such that
\begin{align*}
    \vec{r}=a\vec{x}, \qquad \vec{\nabla}_{x}=a\vec{\nabla}_{r}, \\
    \left(\frac{\partial f(\vec{x},t)}{\partial t}\right)_{r}=\left(\frac{\partial f}{\partial t}\right)_{x}-\frac{\dot{a}}{a}(\vec{x}\cdot\vec{\nabla}_{x})f, \\
    \vec{u}=\dot{a}\vec{x}+\vec{v}(\vec{x},t).
\end{align*}

 Focusing on the scalar modes of the cosmological perturbations that are necessary to calculate the skewness of the matter density field one has to keep nonlinear terms resulting from the product of linear contributions as $(\delta^{\prime})^2, \delta \delta^{\prime}, \delta^2$ and $\delta \,\varphi$. We proceed by isolating the divergence of the velocity field term in the perturbed continuity Eq. (\ref{cont}) and inserting it into  Eq. (\ref{Euler}). With this procedure we find the following equation for the density contrast $\delta=\delta\rho/\rho$
 
\begin{eqnarray}
\label{denscontrpert}
    &\delta''+\delta'\left[ 3(c_s^{2}-w)+\left(1+\frac{\mathcal{H}'}{\mathcal{H}}\right)-\frac{w'}{1+w+(1+c_{s}^{2})\delta} \right]-{\delta'}^{2}\left[ \frac{4/3+c_s^{2}}{1+w+(1+c_{s}^{2})\delta} \right] \nonumber \\
    &+\delta\delta'\left[ \frac{(w-c_s^{2})(5+3c_s^{2})-{c^{2}_{s}}^{\prime}}{1+w+(1+c_{s}^{2})\delta} \right]+\delta^{2}\left[ \frac{3(w-c_s^{2}){c^{2}_{s}}^{\prime}+3(w-c_s^{2})^{2}}{1+w+(1+c_{s}^{2})\delta} \right] \nonumber \\
    &+3\delta\left[ ({c}_s^{2\prime}-w')+\left(1+\frac{\mathcal{H}'}{\mathcal{H}}\right)(c_s^{2}-w)+\frac{w'(w-c_s^{2})}{1+w+(1+c_{s}^{2})\delta}\right] \nonumber \\
    &+\frac{1+w+(1+c_{s}^{2})\delta}{\mathcal{H}^{2}}\nabla^{2}\varphi=\frac{1+w+(1+c_{s}^{2})\delta}{\mathcal{H}^{2}}\nabla\left[\frac{\nabla(c_s^{2}\delta)}{1+w+(1+c_{s}^{2})\delta}\right],
\end{eqnarray}

where the prime represents a derivative with respect to $N=\ln(a)$ and $\mathcal{H}=a^{\prime}/a$.
Eq. (\ref{denscontrpert}) represents the most general equation for the evolution of the density contrast of a fluid with effective sound speed  given by $c^{2}_s=\delta P/\delta\rho$ sourced by a gravitational potential $\varphi$. 

Let us now apply Eq. (\ref{denscontrpert}) to the matter fluid. In the case of a pressureless component both the equation of state parameter and the sound speed  vanish i.e., $w_{m}=c_{s(m)}^{2}=0$. Hence, the contributions proportional to $\delta \delta ^{\prime}$ and $\delta^2$ are eliminated. Therefore Eq. (\ref{denscontrpert}) reduces to the well-known equation

\begin{equation}\label{denscontrmatter}
    \delta_{m}''+\delta_{m}'\left(1+\frac{\mathcal{H}'}{\mathcal{H}}\right)-\delta_{m}^{'2}\left(\frac{4/3}{1+\delta_{m}}\right)+(1+\delta_{m})\frac{\nabla^{2}\varphi}{\mathcal{H}^{2}}=0.
\end{equation}

The contribution of the gravitational potential $\varphi$ is provided by the perturbed Poisson equation. If the universe is filled with both pressureless matter and dark energy, the perturbed Poisson equation reads
\begin{equation}
    \nabla^{2}\varphi=-\frac{3}{2}\Omega_{m}\mathcal{H}^{2}\delta_{m}-\frac{3}{2}\Omega_{de}\mathcal{H}^{2}\left[1+3c_{s(de)}^{2}\right]\delta_{de}.
\end{equation}

Hereafter we use the simplified notation $c_{s(de)}^{2}\equiv c_{de}^{2}$ for the effective DE sound speed. We will treat $c_{de}^{2}$ as a constant free parameter in our analysis. At this point one can notice that all dark energy perturbations effects on the gravitational potential can be eliminated by setting $c_{de}^{2}=-1/3$. However, of course, this choice does not represent a realistic dark energy fluid.

In order to decouple Eq. (\ref{denscontrmatter}) into  first order and second order contributions, we  expand the matter density contrast in terms of the growth functions $D_i$ \cite{Bernardeau:1993qu,Bernardeau:2001qr}
\begin{equation}\label{delext}
    \delta_m=\sum_{i=1}^{\infty}\delta_{i}=\sum_{i=1}^{\infty}\frac{D_{i}(\eta)}{i!}\delta_{0}^{i} \simeq
D_{1}\delta_{0}+\frac{D_{2}}{2}\delta_{0}^{2} +\mathcal{O}(\delta_0^3)\,,
\end{equation}
where $\delta_0$ is the density contrast at some initial epoch.
By neglecting $\mathcal{O}(\delta_0^3)$ terms in the above sum the resulting dynamics is given by two equations for the evolution of the first order ($D_1$) and second order $(D_2)$ growth functions. They read, respectively,
\begin{eqnarray}\label{D1eq}
    D_{1}''+D_{1}'\left(1+\frac{\mathcal{H}'}{\mathcal{H}}\right)-\frac{3}{2}D_{1}\Omega_{m}-\frac{3}{2}\Omega_{de}\delta_{de}(1+3c_{de}^{2})&=&0,
\\
\label{D2eq}
    D_{2}''+{D'}_{2}\left(1+\frac{\mathcal{H}'}{\mathcal{H}}\right)-\frac{8}{3} {D}_{1}^{\prime 2}-\frac{3}{2}D_{2}\Omega_{m}-3D_{1}^{2}\Omega_{m}-3D_{1}\Omega_{de}\delta_{de}(1+3c^{2}_{de})&=&0\,.
\end{eqnarray}

The dark energy perturbation terms can be easily identified in the above equations. They can be switched off by setting $\delta_{de}=0$. Notice that we kept $\delta_{de}$ at first order because dark energy is expected to be much less clustered than  matter.

With these equations we can finally calculate the skewness $S_3$, defined as \cite{Bernardeau:2001qr}
\begin{equation}\label{S3}
    S_{3}=\frac{3D_{2}}{D^{2}_{1}}.
\end{equation}
The skewness is of course just the first non-Gaussian moment, and one could include higher and higher moments as e.g., the kurtosis $S_4=4 (D_3/D_1^3)+12(D_{2}^{2}/D_{1}^{4})$ \cite{Bernardeau:1993qu} to obtain a more complete picture. However, we will see that the skewness depends quite weakly on cosmological parameters, as already found in previous works, and we expect then that higher moments are even less sensitive.

To evaluate $S_3$ in the case the matter density field is sourced by the DE fluctuations $\delta_{de}$ we need now to adapt Eq. (\ref{denscontrpert}) to the dark energy fluid. Since Eqs. (\ref{D1eq}) and (\ref{D2eq}) are sourced by the first order DE density perturbation $\delta_{de}$, we need its linear equation in  Fourier space
\begin{align}\label{DEeq}
    \delta_{de}''&+\delta_{de}'\left[3(c_{de}^{2}-w_{de})+\left(1+\frac{\mathcal{H}'}{\mathcal{H}}\right)-\frac{w_{de}'}{1+w_{de}}\right]\nonumber \\ &+3\delta_{de}\left[(c_{de}^{2\prime}-w_{de}')+\left(1+\frac{\mathcal{H}'}{\mathcal{H}}\right)(c_{de}^{2}-w_{de})-\frac{w_{de}'(c^{2}_{de}-w_{de})}{1+w_{de}}\right]\nonumber \\
    &-(1+w_{de})\frac{3}{2}\left[\Omega_{m}D_{1}+\Omega_{de}(1+3c_{de}^{2})\delta_{de}\right]+\frac{c_{de}^{2} k^2}{\mathcal{H}^{2}}\delta_{de}=0.
\end{align}
When DE perturbations are included, a dependence on the wavenumber $k$ appears in Eq. (\ref{DEeq}). We have checked however that  $S_3$  is very weakly dependent on the $k$-value for the range of scales of cosmological interest, with changes by at most $10^{-5}$. Therefore we neglect the $k$-dependence and adopt for all results  the scale $k=0.01$ $h/$Mpc, close to the range of scales effectively observed in large-scale surveys.

The DE  sound speed $c^{2}_{de}$ will be taken as an additional free parameter.
Hereafter we will adopt the following values

\begin{itemize}
    \item $c^{2}_{de}=0$, a situation in which DE perturbations follow the behavior of the pressureless-like clustering fluid (see e.g. \cite{Creminelli:2009mu});
    \item $c^{2}_{de}=1$, the effective sound speed  of a scalar field component; 
    \item $c^{2}_{de}=1/3$, as indicated recently by the analysis of Ref. \cite{Moss:2021obd}. 
\end{itemize} 
The analysis of \cite{Moss:2021obd} refers to a model different from the one of this work, but suggests the possibility of a dark energy model with a preferred sound speed different from zero or unity,  so we adopt this value for qualitative purposes.
Finally, besides giving accurate fits for these values, we also include $c^2_{de}$ as a third parameter along with $\Omega_{m0}$ and $w_{de}$.
We also consider for comparison the case in which no DE perturbations are present ($\delta_{de}=0$), corresponding to $c_{de}\to\infty$. This case is however hardly distinguishable from $c_{de}=1$.

\section{Results and numerical fits}

An expansion for the skewness value around the  Einstein-De-Sitter universe ($\Omega_{m}=1$ and $\Omega_{\Lambda}=0$) is often employed in the literature. A widely known result is provided by Bernardeau et al. (2002) \cite{Bernardeau:2001qr} where a weak dependence of the skewness on the $\Omega_{m0}$ parameter $S_{3}=34/7+6/7(\Omega^{-0.03}_{m0}-1)$  has been found for the case where $\Lambda=0$.
The main goal of this paper is to extend this analysis by including the effects of dark energy fluctuations as a function of the matter density parameter $\Omega_{m0}$,  the dark energy equation of state  $w_{de}$, and its squared sound speed $c^2_{de}$.  

\begin{figure}
\centering
\includegraphics[width=0.65\textwidth]{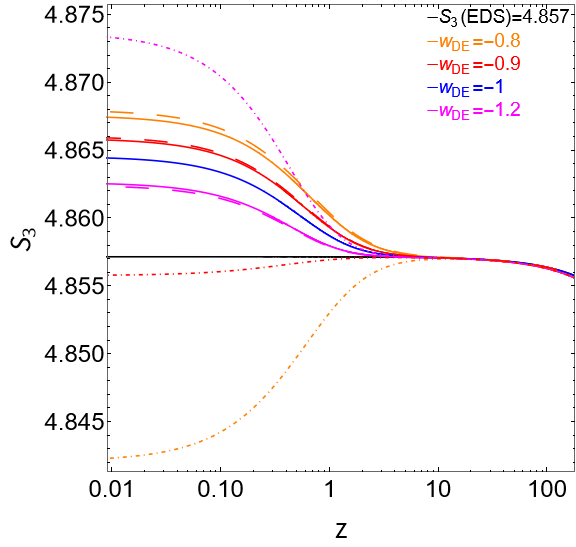}
\caption{Skewness evolution as a function of redshift. Solid lines represents $c^2_{de}=1$ cases, dot-dashed $c^2_{de}=0$ and dashed no DE perturbations. The colors represents the following cases: blue for a $\Lambda$CDM case i.e., $\Omega_{m0}=0.3$, $w_{de}=-1$ (cases for $w_{de}=-1$ and $c^2_{de}=0$, $c^2_{de}=1$ and no DE perturbation are indistinguishable); black for EdS case i.e., $\Omega_{m0}=1$. Then, always putting $\Omega_{m0}=0.3$, we employ orange for $w_{de}=-0.8$; red for $w_{de}=-0.9$ and  magenta for $w_{de}=-1.2$.} 
\label{FigSkewScaleFactor}
\end{figure}

We begin our analysis by showing in Fig. \ref{FigSkewScaleFactor} how the skewness evolves in time for a $\Lambda$CDM cosmology with parameters $\Omega_{m0}=0.3, w_{de}=-1$ (blue line) and for the Einstein-de Sitter (EdS) model (black line). As expected, the non-Gaussianity increases with time reaching asymptotically a constant value for both cases. One can clearly see the manifestation of dark energy around  redshift 1. We also show models with dark energy perturbations for  $w_{de}=-0.8$, $w_{de}=-0.9$ and $w_{de}=-1.2$. We extend our analysis comparing different cases of $c^{2}_{de}$ and the case without dark energy perturbations i.e., setting $\delta_{de}=0$. One can notice that for $c^{2}_{de}=0$ the skewness is more sensitive than the other cases to the equation of state.

\begin{figure}
\centering
\includegraphics[width=0.44\textwidth]{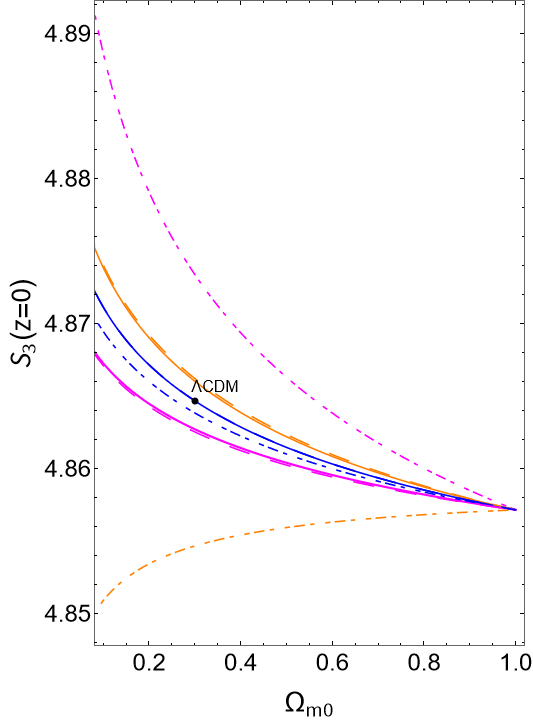}
\includegraphics[width=0.45\textwidth]{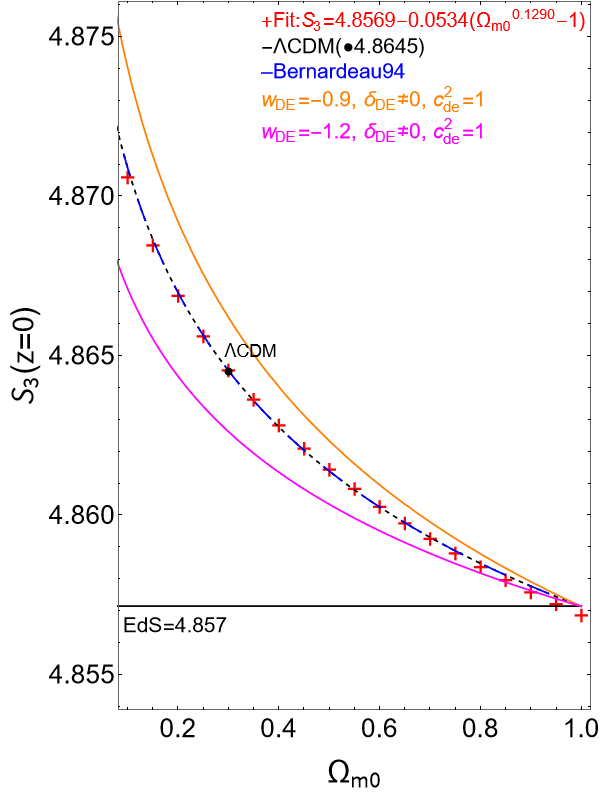}
\caption{Dependence of the skewness on the  density parameter $\Omega_{m0}$. In the left panel the dot-dashed lines represent $c^{2}_{de}=0$, solid lines $c^{2}_{de}=1$ and dashed lines is the case without dark energy perturbations ($\delta_{de}=0$). The colors refer to different $w_{de}$ values: blue for $w_{de}=-1$, magenta for $w_{de}=-1.2$ and orange for $w_{de}=-0.9$. The dashed and solid blue lines are indistinguishable. On the right side we show the skewness for  $\Lambda \neq 0$ by Bernardeau (1994) \cite{Bernardeau:1993qu} (dashed blue) compared to our result (dashed black) for the $\Lambda$CDM case ($w_{de}=-1$, $\Omega_{m0}=0.3$ (black dot) and no dark energy perturbations $\delta_{de} = 0$). The black line is for the EdS skewness value of $34/7 = 4.857$. The fit Eq.(\ref{FIT1}) is represented by the red cross symbols.}
\label{FigSkewness}
\end{figure}

We solve now  Eq. (\ref{D1eq}), Eq. (\ref{D2eq}), and Eq. (\ref{DEeq}) varying  systematically the parameters $\Omega_{m0}$ and $w_0$. We choose $a_{i}=1/(1 + z_{i})$, being $z_{i}\simeq 300000$ so that when $t\rightarrow 0$ we have a EdS cosmology \cite{Bernardeau:1993qu},  and we assume as initial conditions $D_{1}(a_{i})\propto a$ and $D_{2}(a_{i})\propto a^{2}$. We show  in  Fig. \ref{FigSkewness}  today's skewness value $S_3$ as a function of  $\Omega_{m0}$  for various values of $w_{de}$ and $c^{2}_{de}$. On the left plot of Fig.\ref{FigSkewness} we notice again that for $c^{2}_{de}=0$ the skewness depends more sensitively on the cosmological parameters. In fact, we see that dark energy  with $c^{2}_{de}=0$  increases (for phantom cases \cite{Caldwell:1999ew}) or decreases (for quintessence \cite{Benabed:2001dm}) today's skewness value. This indicates that the skewness can be a useful tool to test models with a vanishing sound speed $c^{2}_{de}=0$.
Overall, however, we see that the typical change in $S_3$ from model to model is small,  at most 1\%. 

In order to provide an expression for $S_3$ as a function of the cosmological parameters, we have performed phenomenological fits to the numerical results. To assess the quality of the fits, we adopt two measures, the maximum relative deviation (MRD) and the average standard relative deviation (ASRD), defined respectively as the following percentages
\begin{align}
    {\rm MRD}=& 100\, {\rm Max}\Big{\lvert}\frac{{\rm Fit}(i)-{\rm Data}(i)}{{\rm Fit}(i)}\Big{\rvert}, \nonumber \\
    {\rm ASRD}=&100\sqrt{\frac{1}{N}\sum\Big{(} \frac{{\rm Fit}(i)- {\rm Data}(i)}{{\rm Fit}(i)}\Big{)^{2}}},
\end{align}
where ${\rm Data}(i)$ represents the numerical outcomes of our system of equations for a given set parameter values $i$ and ${\rm Fit}(i)$ is the fitted function. We aim at an average (ASRD) and maximum (MRD) precision much better than 1\% across the entire parametric range.

In Fig. \ref{FigSkewness} (right side) we show that the skewness calculated for a $\Lambda$CDM ($\delta_{de}=0$) case is almost identical to Bernadeau's results in \cite{Bernardeau:1993qu} for  $\Lambda\neq 0$. A fit to our results is  provided as the following equation:
\begin{equation}
S_3(\Omega_{m0}) = 4.857 - 0.053(\Omega_{m0}^{0.129}-1).
\label{FIT1}
\end{equation}
For this fit we obtain the following values quality measurements: MRD=$0.01\%$ and ASRD$=0.002\%$.
We also plot in the same figure  the cases for dynamical dark energy models for quintessence $w_{de}=-0.9$  and phantom $w_{de}=-1.2$ cosmologies including dark energy perturbations, both  with $c_{de}^{2}=1$ as in the left panel.

\begin{figure}
\centering
\includegraphics[width=0.5\textwidth]{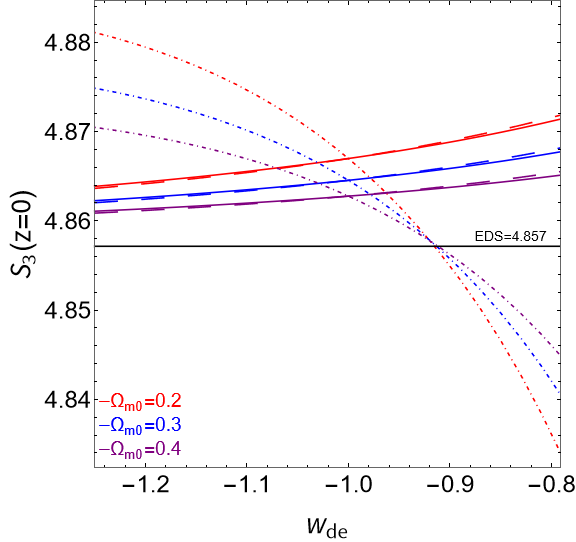}
\caption{Dependence of $S_3$ on the dark energy equation of state $w_{de}$. The colored curves refers to different values of the matter density parameter, namely $\Omega_{m0}=0.2$ (red), $\Omega_{m0}=0.3$ (blue) and $\Omega_{m0}=0.4$ (purple). The black line sets the Einstein-de Sitter value $34/7$. We assumed $c^{2}_{de}=1$ for solid lines, $c^{2}_{de}=0$ for dot-dashed lines  and no DE perturbations ($\delta_{de}=0$) for dashed lines (hardly visible next to the solid curves).}
\label{FigPhanQuin}
\end{figure}

Our goal now is to obtain a general fitting formula for $S_3$ that includes the dependence on both  $\Omega_{m0}$ and $w_{de}$ for various values of $c^{2}_{de}$. 
We propose the following parametrization, defined by six  constants $a,b,..f$:
\begin{equation}\label{Fiteq}
    S_{3}(\Omega_{m0},w_{de})=a+b(\Omega_{m0}^{c}-1)+d|(1+w_{de})|^{e}+f(1+\Omega_{m0}w_{de}).
\end{equation}
By setting $d=e=f=0$ one obtains the previous formula  (\ref{FIT1}). Also, the constant $S_3$ value of the EdS model is immediately retrieved in the limit $\Omega_{m0}=1$ and $w_{de}=-1$.

Tables \ref{tablephantom} (phantom regime only) and \ref{tablequint} (quintessence regime only) report the values for the  fitting free parameters $ a, b, ...,f$  and their respective MRD and ASRD values  for some values of $c_{de}^{2}$.

\begin{figure}
\centering
\includegraphics[width=0.43\textwidth]{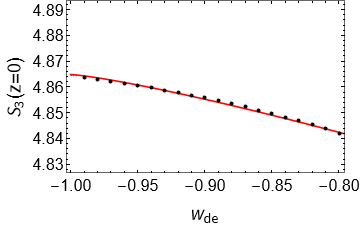}
\includegraphics[width=0.43\textwidth]{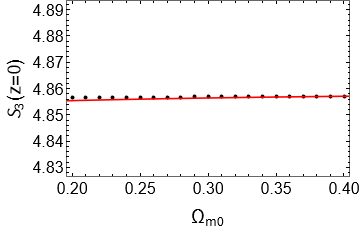}
\includegraphics[width=0.43\textwidth]{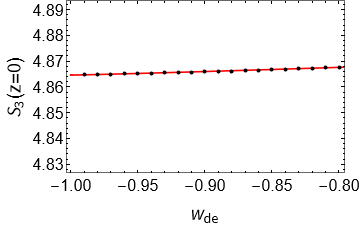}
\includegraphics[width=0.43\textwidth]{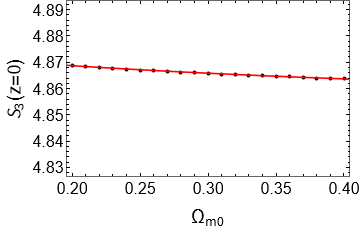}
\caption{Variation of skewness with $\Omega_{m0}$ and $w_{de}$. On the left panels we fix $\Omega_{m0}=0.3$; on the right panels we fix  $w_{de}=-0.9$. The red line represents the fit from Eq. (\ref{Fiteq}) and Table \ref{tablequint},  and the black dots the data. Here we have the case for $c_{de}^{2}=0$ (top row) and $c_{de}^{2}=1$ (lower row).}
\label{Figfitquin}
\end{figure}

In order to visualise the accuracy of the fitting formula Eq.(\ref{Fiteq}) with the parameters of Table(\ref{tablequint}), we show in  Fig.\ref{Figfitquin} the comparison between the numerically generated data  and the fit. Here we have chosen to display the quintessence case; since this is the case with largest MRD and ASRD values, all other parameter configurations produce a better fit.

\begin{table}[]
\begin{tabular}{|c|c|c|c|c|c|c|c|c|}
\hline
Phantom ($-1.25\leq w_{de}\leq-1$) & a     & b      & c     & d      & e     & f      & MRD\% & ASRD\% \\ \hline
$c_{de}^{2}=0$                     & 4.858 & 0.049  & 1.492 & 0.052  & 0.841 & 0.069  & 0.04  & 0.01   \\ \hline
$c_{de}^{2}=1$                     & 4.857 & -0.034 & 0.625 & -0.012 & 0.926 & -0.015 & 0.01  & 0.002  \\ \hline
$c_{de}^{2}=1/3$                   & 4.857 & -0.034 & 0.621 & -0.012 & 0.930 & -0.015 & 0.01  & 0.002  \\ \hline
\end{tabular}
\caption{Fit values for Eq. (\ref{Fiteq}) for the phantom regime  $-1.25\leq w_{de}\leq-1$.}\label{tablephantom}
\end{table}

\begin{table}[]
\begin{tabular}{|c|c|c|c|c|c|c|c|c|}
\hline
Quintessence ($-1\leq w_{de}\leq-0.8$) & a     & b      & c     & d      & e     & f      & MRD\% & ASRD\% \\ \hline
$c_{de}^{2}=0$                         & 4.853 & 0.186  & 0.946 & -0.216 & 1.143 & 0.198  & 0.1   & 0.01   \\ \hline
$c_{de}^{2}=1$                         & 4.858 & -0.050 & 0.574 & 0.026  & 1.075 & -0.026 & 0.02  & 0.002  \\ \hline
$c_{de}^{2}=1/3$                       & 4.858 & -0.050 & 0.556 & 0.024  & 1.070 & -0.025 & 0.02  & 0.002  \\ \hline
\end{tabular}
\caption{Fit values for Eq. (\ref{Fiteq}) for the quintessence regime  $-1\leq w_{de}\leq-0.8$.}\label{tablequint}
\end{table}

\begin{table}[]
\begin{tabular}{|c|c|c|c|c|c|c|c|c|c|}
\hline
                                       & a     & b      & c     & d       & e      & f      & g      & MRD\% & ASRD\% \\ \hline
Phantom ($-1.25\leq w_{de}\leq-1$)     & 4.856 & -0.032 & 0.413 & -0.008  & 1.012  & -0.007 & -0.002 & 0.3   & 0.03   \\ \hline
Quintessence ($-1\leq w_{de}\leq-0.8$) & 4.859 & -0.053 & 0.172 & -158780 & 14.258 & 0.003  & 0.003  & 0.8   & 0.06   \\ \hline
\end{tabular}
\caption{Fit values for Eq. (\ref{Fitcde}) for the phantom and quintessence regimes.}\label{tablecde}
\end{table}

\begin{figure}
\centering
\includegraphics[width=0.43\textwidth]{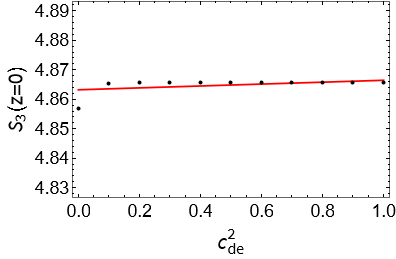}
\caption{Variation of skewness with $c^2_{de}$. The red line represents the fit Eq.(\ref{Fitcde}) and Table \ref{tablecde} for $\Omega_{m0}=0.3$ and $w_{de}=-0.9$; the black dots are the data.}
\label{Figfitcde}
\end{figure}

We also propose a general fitting formula with $c_{de}^{2}$ as a variable \cite{Sergijenko:2014pwa,Basse:2010qp,Creminelli:2009mu}. The results are presented in Table(\ref{tablecde}) and Fig.(\ref{Figfitcde}), where its important to notice that the fit is not very good for $c_{de}\to 0$ (in which case one should use Tables \ref{tablephantom} and \ref{tablequint}):
\begin{equation}\label{Fitcde}
    S_{3}(\Omega_{m0},w_{de},c_{de}^{2})=a+b(\Omega_{m0}^{c}-1)+d|(1+w_{de})|^{e}+f(1+\Omega_{m0}w_{de})+g(c_{de}^{2}+w_{de}).
\end{equation}

We can see in both Table (\ref{tablephantom}) and Table (\ref{tablequint}) that the fitted values for the cases $c^{2}_{de}=1/3$ and $c^{2}_{de}=1$ are quite similar, see also in Fig.(\ref{Figfitcde}). Indeed, the case $c^{2}_{de}=0$ is distinct from any $c^{2}_{de}\neq0$ case. The reason is of course that only a vanishingly small $c_{de}$ can damp the strong oscillations induced by  the $(k/{\cal H})^2\approx 10^3$ term in Eq. (\ref{DEeq}). This behavior can also be seen in the previous plots Fig.(\ref{FigSkewness}) and Fig.(\ref{FigPhanQuin}), where $c^{2}_{de}=0$ is the case  more sensitive to cosmological parameters. As soon as $c_{de}>{\cal H}/k$, the skewness is almost independent of $c_s$ and very  weakly dependent on cosmological parameters.

Apart from the theoretical analysis performed so far, it is also important to have contact with observed quantities. 
However, the value of $S_3$ for biased tracers is actually expected to be shifted by a factor and a constant that depend on the linear and non-linear bias parameters \cite{1993ApJ...413..447F}, plus a correction that depends on the cell window function. One could compare the data with $S_3$ and find the best fit values of the bias parameters, but here we limit ourselves to plotting $S_3$ against the data to have a general idea of the trend.

\begin{figure}
\centering
\includegraphics[width=0.5\textwidth]{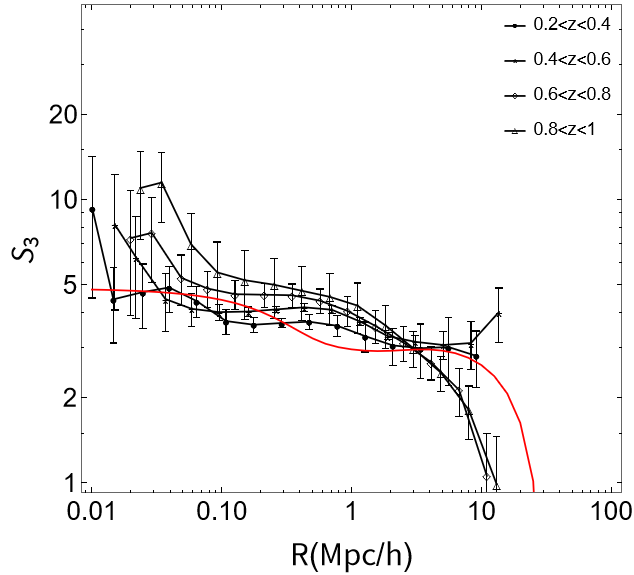}
\caption{Skewness  compared to observational data from \cite{Wolk:2013db}. The red line represents the theoretical plot using our set of equations with $\Omega_{m0}=0.27$ and $H_{0}=71$ km/s/Mpc that was also used in \cite{Wolk:2013db}. The cases for different values of $w_{de}$ and $c^{2}_{de}$ are indistinguishable in this scale. Also the theoretical curve is almost independent of $z$; we plot here the $z=0.9$ case.}
\label{ObservPlot}
\end{figure}

In Fig. \ref{ObservPlot} we show our results against the observed values from Ref. \cite{Wolk:2013db} from the CFHTLS-Wide survey. In this analysis the skewness is calculated using a counts-in-cells technique for several intervals of redshift. Here we managed to reproduce a larger range of theoretical results than in \cite{Wolk:2013db} using a well-behaved (i.e., less oscillating than the top-hat) window function, namely  
\begin{equation}
    W({ k R})=\frac{\sqrt{2} e^{-\frac{{ k R}}{\sqrt{2}}} }{{ k R}}\sin \left(\frac{{ kR}}{\sqrt{2}}\right),
\end{equation} 
where $W(kR)$ is the Fourier transform of the  smoothed top-hat spherically symmetric region with characteristic radius
$R$ defined as $W(r)=1/(1+(r/R)^4)$
normalized to unity. Then, the following expression is used \cite{Wolk:2013db,Bernardeau:2001qr,Bernardeau:1994vz,Juszkiewicz:1993uw}
\begin{equation}
    S_{3}=\frac{34}{7}+ \gamma_{1}.
\end{equation}
where $\gamma_{1}=d \ln \sigma^{2}(R)/d \ln R$ is the count-in-cells correction  \cite{Bernardeau:1993qu,Wolk:2013db} and   the variance of the density field in real space is given by 
\begin{equation}
    \sigma^2(R)=\frac{1}{2\pi^2}\int P(k,z)\, W^2(kR)\,k^2\,dk.
\end{equation}
where $P(k,z)$ is a $\Lambda$CDM power spectrum at redshift $z$. We can see that the general trend is consistent with the data, but clearly the precision is still far from allowing a robust constraint on the cosmological and the bias parameters.

\section{Conclusions}

This work presents numerical estimates of $S_3$ (the third moment of the matter density field) when including dark energy perturbations while varying $\Omega_{m0},w_{de}$ and $c^2_{de}$. Whereas a cosmological constant affects the matter clustering features only via the background expansion, a dynamical dark energy field with its associated perturbations can lead to noticeable imprints on the large scale structure. This means that a measure of $S_3$ within a precision better that 1\% can put new constraints on cosmological parameters. How well this goal can be achieved with future surveys remains to be seen.

In a previous work, \cite{Velten:2019edo}, the universe content was modeled as a effective single fluid (an admixture of dark energy and matter), in which  the dark energy contributions have been considered via the total (the single fluid realization) equation of state parameter and sound speed. In contrast, in this work  we consider the universe as composed by two separated fluids, both of which being perturbed.  As expected, we obtain now  values for  today's $S_3$ much closer to the $\Lambda$CDM case (a factor of three smaller than in \cite{Velten:2019edo}), in  agreement therefore with standard results presented in the literature, e.g. \cite{Bernardeau:1993qu,2001ApJ...548...47G, Wolk:2013db}.

We found a higher sensitivity to the cosmological parameters when the  DE sound speed velocity   $c^{2}_{de}$ is set to $0$.
Depending on the regime being phantom or quintessence, a vanishing sound speed  can increase or lower the skewness in a significant way. On the other hand, if the dark energy background equation of state is set to $w_{de}=-1$ the skewness can not be usefull to distinguish the perturbative properties of the dark energy component as e.g., its speed of sound (see Fig.(\ref{FigPhanQuin})).

We provided several fitting formulas accurate to within (or better than) 0.1\% on average for $S_3$ as a function of the cosmological parameters  $\Omega_{m0}, w_{de}$ and $c^2_{de}$. These fits may be useful to compare the cosmological models to large scale structure at higher orders. 

In future work, we plan to extend this analysis to include more cosmological parameters and  modified gravitational models, as e.g. in \cite{2004PhRvL..92r1102A,Tatekawa:2008bw}.

\noindent
\section*{Acknowledgments}
 Raquel Fazolo thanks  CNPq and FAPES for financial support. Hermano Velten thanks CAPES/CNPq and Proppi/UFOP for partial finantial support. LA acknowledges financial support from DFG project  456622116.


\end{document}